\documentclass[12pt,preprint]{aastex}

\usepackage{amsfonts}
\usepackage{amsmath}
\usepackage{amssymb}
\usepackage{pdflscape}
\usepackage{rotating}
\usepackage{txfonts}
\usepackage{longtable}
\usepackage{xspace}
\usepackage{ifpdf}
\usepackage{eurosym}
\usepackage{rotating}
\usepackage{hyperref}

\def\cygx1{Cyg X-1}

\def\ss{}
\def\ssss{}
\def\saim{}

\input psfig.sty

\shorttitle{PSF in gamma-ray telescopes}
\shortauthors{SS}

\begin{document}

\title{On the Angular Resolution of the AGILE gamma-ray imaging detector}
\author{Sabatini S., Donnarumma I., Tavani M., Trois A., Bulgarelli A., 
Argan A., Barbiellini G., Cattaneo P. W., Chen A., Del Monte E., Fioretti V., 
Gianotti F., Giuliani A.,  Longo F., Lucarelli F., Morselli A.,
 Pittori C., Verrecchia F., Caraveo P.}

\begin{abstract}
We present a study of the Angular Resolution  of the {\it
AGILE} gamma-ray imaging detector (GRID) that is operational in
space since April 2007.
The AGILE instrument is made of an array of 12 planes
each equipped with a Tungsten converter and Silicon microstrip
detectors and is sensitive in the
energy range 50 MeV - 10 GeV. 
Among the space instruments devoted to gamma-ray
astrophysics, {\it AGILE} uniquely exploits an analog readout
system with dedicated electronics coupled with Silicon detectors.
We show the results of Monte Carlo simulations carried out to
reproduce the gamma-ray detection by the {\it GRID}, and we
compare them to in-flight data. {\ss We use the Crab (pulsar $+$ Nebula)
system for discussion of real data performance,
 since its $E^{-2}$ energy spectrum is representative
of the majority of gamma-ray
sources. For Crab-like spectrum
sources, the {\it GRID} angular resolution
(FWHM of $\sim \rm4^{\rm \circ}$ at 100 MeV; $\sim0.8^{\rm \circ}$ at 1 GeV; 
$\sim 0.9^{\rm \circ}$ integrating the full energy 
band from 100 MeV to tens of GeV) is 
stable across a large field of view, being characterized by a 
flat response up to 30$^{\rm \circ}$ off-axis. A comparison of the 
angular resolution obtained by the two operational 
gamma-ray instruments, {\it AGILE-GRID} and {\it Fermi-LAT}, 
is interesting in view of future gamma-ray missions, that are 
currently under study. The two instruments exploit different detector 
configurations affecting the angular resolution: the former being
optimized in the readout and track reconstruction especially in the low-energy band, 
the latter in terms of converter thickness {\ssss and power consumption}. 
We show that, despite these differences, the angular resolution 
of both instruments is very similar between 100 MeV and a
few GeV.}

\end{abstract}


\section{Introduction}\label{sec:intro}

Gamma-ray astrophysics in space enormously advanced since the
first detection of photons above 100 MeV \citep{kraushaar72}. The
progression of space missions and instruments over the last
decades (OSO-3, SAS-2, COS-B, EGRET on board of the Compton
Gamma-Ray Observatory) led to  improvements of the overall
detector performance, {\ss in terms of both angular resolution and
sensitivity. In addition to this, the progressively wider Field-of-View 
(FoV) allowed a continuous monitoring of the variable gamma-ray sky.}
Following the early cosmic gamma-ray detection by the OSO-3
satellite \citep{kraushaar72}, the first gamma-ray telescope SAS-2
launched in 1972 reached an angular resolution of a few degrees
\citep{fichtel75}. This mission was followed by the European
mission COS-B, launched in 1975 August 8 and operational for 7
years \citep{mayer-1979,swanenburg-1981}. The Compton Gamma-Ray
Observatory (CGRO), active between 1991-2000, provided the first
complete investigation of the gamma-ray sky. In particular, CGRO
hosted the Energetic Gamma-Ray Experiment Telescope (EGRET),
operating in the energy range 30 MeV - 30 GeV
\cite{fichtel97, thompson93, thompson98}. {\ss 
Pre-{\it AGILE}/Pre-{\it Fermi} space instruments were mostly based on spark chamber
technology.}

{\saim Further improvements to the overall
performance of gamma-ray detectors in space became possible with 
the advent of solid-state Silicon
detector technology.} The scientific objectives of the new generation of instruments 
required the following enhancements: (1) improving the gamma-ray angular resolution near 100
MeV by at least a factor of 2-3 compared to EGRET; (2) obtaining
the largest possible FoV at 100 MeV reaching 2.5-3
sr ; (3) increasing flux sensitivity near 100 MeV.
\noindent The current generation of gamma-ray space instruments, {\it
AGILE-GRID}  (Gamma-Ray Imaging Detector) and {\it Fermi-LAT}
(Large Area Telescope), launched in April 2007 and June
2008 respectively, were designed to achieve these objectives.
Both instruments
 make use of tungsten-silicon detectors for the
conversion and detection of gamma-ray photons, {\ss having a common overlapping
energy range in the $\sim50$ MeV - 10 GeV band} 
\citep{tavani09, atwood09}. 
The gamma-ray
detector is structured to form a ``Silicon Tracker''  made of
several trays, each containing a tungsten layer (used as
converter of the incident $\gamma$-ray photon into an
e$^{+}$/e$^{-}$ pair) and two silicon strip layers (used to track
the the e$^{+}$/e$^{-}$ pair path across the instrument, through
their (x,y) projections along each tray). The detected tracks are
identified and fitted by a Kalman filter that allows for the
reconstruction of the original direction and energy of the
incident photon.  The main source of contamination in the
detection of  $\gamma$-ray photons is then by charged particle
tracks, which can be confused with e$^{+}$/e$^{-}$ pairs. For this
reason, background rejection filters
 are used both on-board and on-ground, to process the data and
obtain a final discrimination and classification of the events.\\

An important parameter in the assessment of the overall instrument
scientific performance is the resulting angular resolution, {\ss i.e. the 
minimum distance at which two close sources are distinguished as
separated. The Point Spread Function (PSF) describes the response of an 
imaging instrument to a point source and the angular resolution {\saim is
usually} described by either the Full Width at Half Maximum (FWHM) of the 
PSF radial profile or the 
68$\%$ containment radius (CR$_{68\% }$) of the PSF.} Key parameters of
the instrument configuration affecting the angular resolution of $\gamma$-ray
{\saim solid-state} silicon detectors are:
\begin{itemize}
\item {\ss 
the size of the silicon strips (the pitch) and the silicon 
detector readout system (e.g., either digital or analog) 
resulting in different
effective spatial resolution of the particle trajectory due to
charge coupling between adjacent silicon strips; 
\item the distance between consecutive silicon planes
that, combined with the spatial resolution, defines the {\saim limiting} angular resolution};
 \item the thickness of tungsten layers promoting gamma-ray photon conversion,
 and at the same time inducing multiple scattering that leads to a  
degradation of the charged particle tracks;
\item {\ssss the reconstruction and event classification algorithms.} 
 
\end{itemize}

Table A1 in the Appendix summarizes the main characteristics of
the {\it AGILE} and {\it Fermi} gamma-ray detectors, whose configurations {\saim 
are different} 
in several ways: the former being
optimized in the readout and track reconstruction especially in the low energy band, 
the latter in terms of converter thickness, geometrical area, {\saim and
power consumption}. An
important difference between the two instruments, that can
be {\saim crucial for the }
scientific
performance of the instrument, is the readout system (analog for the {\it
AGILE-GRID} and digital for {\it Fermi-LAT}).


In the following sections, we focus on the characterization of the  {\it
AGILE-GRID} angular resolution both from simulations and from in-flight data. 
In section \ref{sec:angres} {\saim we define }the parameters that are used across the paper 
to describe the instrument angular resolution; section 
\ref{sec:simsetup} describes the Monte Carlo (MC) simulations setup and data 
processing pipeline; section \ref{sec:datasim} shows the results from the
MC simulations; section \ref{sec:datacrab} compares the results of the simulation 
with the in-flight data angular resolution. This paper complements the work
by \cite{chen13}:  
{\saim here }we focus on
simulations characterizing the overall performance of the  {\it
GRID} instrument and in a special study of Crab-like sources, reproducing the
behaviour of the majority of detected cosmic gamma-ray sources;
we also present a direct comparison of  {\it AGILE} and {\it Fermi}
in-flight data for the Crab\footnote{The
case of the Vela pulsar has been addressed by \cite{chen13} and
\citep{ackermann13}.}, and discuss the results concerning
the angular resolution in terms of the different instrument 
configurations (section  \ref{sec:agilevsfermi}), which can be crucial 
in the study of future missions.

\section{The Angular Resolution}\label{sec:angres}

For an imaging telescope, the response in terms of reconstruncted positions 
of a set of photons from a point-like source in the sky defines its Point Spread
Function (PSF). Since this function {\ss can be considered azymuthally 
symmetric {\saim in $\gamma$-ray telescopes} for incidence angle within 30$^{\rm \circ}$ as used in this paper}, 
the effect of 
the dispersion can be described as a function of one parameter, the angular 
distance $\alpha$ between reconstructed and nominal direction. 
Defining the PSF radial profile, P($\alpha$), 
as the probability distribution per steradian of measuring an incoming photon
 at a given angular distance 
$\alpha$ from its true direction, we have therefore:\\ 
\begin{center}
PSF($\alpha$)$d\alpha$=2$\pi$sin($\alpha$)$P(\alpha$)$d\alpha$.
\end{center} 

In the following, we adopt two parameters widely used in the literature 
to {\ss assess the angular resolution in terms of the ``width'' of the 
PSF} \cite[e.g.][]{thompson93}: 
\begin{itemize}
    \item the Full Width at Half Maximum (FWHM) of the probability 
           distribution per steradian, $P(\alpha$)
    \item the 68\% Containment Radius (CR$_{68\%}$) of the probability distribution, PSF($\alpha$). 
\end{itemize}

{\saim The CR$_{68\%}$ is strongly 
related to the source image compactness, taking into account 
the whole contribution of source profile and being more affected by possible 
extended tails; while the FWHM is mainly determined by the central core emission of the source.}

\section{Setup and data processing of the {\it AGILE} simulated data}\label{sec:simsetup}

In order to fully understand how the instrument configuration and
the data analysis pipeline affect the final angular resolution of a gamma-ray
telescope, we carried out dedicated MC simulations of the
{\it AGILE-GRID} analyzing  simulated data with the same pipeline
used {\saim to process} in-flight data. {\ss The simulations were carried out using an 
available simulation tool \cite[``GAMS'', GEANT AGILE
MC Simulator;][]{cocco02, longo02}, implemented during the
 development phase of the {\it AGILE} mission.
The tool makes use of the GEANT3 environment\footnote{
http://wwwasd.web.cern.ch/wwwasd/geant/index.html}
and takes into account all the main components of the  
{\it AGILE} instrument configuration: the spacecraft (bus) MITA 
and the {\it AGILE} payload,
consisting of the CsI Mini-Calorimeter, the silicon-tungsten
Tracker, the Anti-Coincidence system, the X-ray detector ({\it
Super-AGILE}), the thermal shield,
 the mechanical structure and the lateral electronics boards.  
The tool allows to simulate a parallel front of given direction 
for both charged particles and/or photons and its interactions across
the instruments. The spectral energy distribution of the front can
be monochromatic or with any given law.} 
The simulated tracks are then processed by the
DHSIM, the Data Handling SIMulator \cite{argan04,argan08}, which
implements the on-board algorithms for a first track reconstruction,
event classification and background rejection
\citep{giuliani06}. The DHSIM output is a data file in the same
format as the AGILE in-flight telemetry and can be processed 
using the same pipeline as for real data analysis. 
Further background rejection techniques are applied to the data on ground,
producing the final classification of the events. {\ss The
 currently used filter for the scientific analysis of the {\it
AGILE-GRID} data (``FM3.119'') was used in this paper also for the
processing of the simulated data.\\
Each simulation contains $2\times10^{5}$ photons that cross 
the tracker and the data are analyzed with the same pipeline as for real data. 
Typical efficiency after background rejection within 30$^{\rm \circ}$ off-axis angle 
is $\sim 30\%$, leaving therefore a set of $\sim 5\times10^{4}$ events classified 
as photons and used for our calculations in the paper.}

\section{The Angular Resolution for simulated {\it AGILE-GRID} data}\label{sec:datasim}

With the aim of investigating the dependence of the {\saim telescope} angular resolution 
upon the energy of incoming photons and their incident direction in the FoV, 
we carried out simulations of parallel fronts of photons by varying 
those parameters.\\ 
\noindent We describe the incident direction as the composition of a
zenith angle $\theta$ (i.e. the angular distance of the
incoming photon direction from the vertical axis of the Tracker,
also named off-axis angle) and an azimuthal angle $\phi$ (i.e.,
the angular distance between a given axis in the instrument plane
and the projection of the incoming direction {\saim in the plane}). Since the detector
response is mostly  azimuthally symmetric, we expect the main dependence
to be upon $\theta$. 
Simulations were carried out for the following values of these
angles: $\theta = (1^{\rm \circ}, 30^{\rm \circ}, 50^{\rm \circ})$\footnote{{\saim For the on-axis case we use a value of 
$\theta=1^{\rm \circ}$, 
since the case of $\theta=0^{\rm \circ}$ induces a singularity in the software that generates the parallel front of
photons. For our purposes $\theta=1^{\rm \circ}$ well resembles the on-axis case.}}; 
$\phi = (0^{\rm \circ}, 45^{\rm \circ})$. Two different
sets of simulations were performed:

\begin{itemize}

\item parallel fronts of monochromatic photons of energies 50, 100, 200, 400, 1000, 5000
MeV; 
\item parallel fronts of photons with a Crab-like photon spectrum of the
type $E^{-2.1}$, in the energy range 30 MeV - 50 GeV.
\end{itemize}

The simulated data are analysed with the same pipeline as for the real data.
In particular data were analysed using the most recent pipeline 
for the data processing (``BUILD21'') and instrument response functions
(``I0023''). 

\begin{table}[!h] 
\centerline{\bf AGILE-GRID PSF HWHM}
\begin{center}
 \vspace{0.2cm}
\small 
\begin{tabular}{|c|c|c|c|c|c|c|}
  \hline
Centroid  & Energy  & $\theta=1^{\rm \circ}$ & $\theta=1^{\rm \circ}$ & 
$\theta=30^{\rm \circ}$ & $\theta=30^{\rm \circ}$ & Err\\
 Energy  & Band & $\phi=0^{\rm \circ}$  &$\phi=45^{\rm \circ}$  & $\phi=0^{\rm \circ}$ &  $\phi=45^{\rm \circ}$ &      \\
   (MeV) & (MeV) & (deg)  & (deg)  &  (deg)  &  (deg) &   (deg)  \\[0.5ex]
 

  \hline
    50   & 30 - 70      & 4.0 & 4.5 & 4.5 & 4.5 & 0.5  \\   
   100   & 70 - 140     & 2.25 & 2.25 & 2.5 & 2.25 & 0.25 \\   
   200   & 140 - 300    & 1.1 & 1.2 & 1.3 & 1.2 & 0.1  \\   
   400   & 300 - 700    & 0.7 & 0.7 & 0.7 & 0.7 & 0.1  \\   
   1000  & 700 - 1700   & 0.4 & 0.4 & 0.4 & 0.4 & 0.1   \\   
   5000  & 1700 - 10000 & 0.2 & 0.2 & 0.2 & 0.2 & 0.1  \\   
\hline
\end{tabular} 
 \caption{{\small Results from the simulations of monochromatic photon 
parallel beams. Half Width at Half Maximum of the PSF radial 
profile in degrees. Events for the different energy channels are selected 
on the basis of the reconstructed energy, as for real data.
}}\label{tab:monotabfwhm}
\end{center}
\hfill

\centerline{\bf AGILE-GRID PSF CR$_{68\%}$}
\begin{center}
 \vspace{0.2cm}
\small 
\begin{tabular}{|c|c|c|c|c|c|c|}
  \hline
 
Centroid & Energy  & $\theta=1^{\rm \circ}$ & $\theta=1^{\rm \circ}$ & 
$\theta=30^{\rm \circ}$ & $\theta=30^{\rm \circ}$ & Err\\
  Energy  &  Band& $\phi=0^{\rm \circ}$  &$\phi=45^{\rm \circ}$  & $\phi=0^{\rm \circ}$ &  $\phi=45^{\rm \circ}$ &      \\
   (MeV) & (MeV) & (deg)  & (deg)  &  (deg)  &  (deg) &   (deg)  \\[0.5ex]
 \hline
    50   & 30 - 70        & 7.5 & 7.5 & 8.5 & 8.5 & 0.5  \\
   100   & 70 - 140       & 4.3 & 4.3 & 4.7 & 5.0 & 0.25 \\
   200   & 140 - 300      & 2.2 & 2.2 & 2.6 & 2.7 & 0.1 \\
   400   & 300 - 700      & 1.2 & 1.2 & 1.3 & 1.4 & 0.1 \\
   1000  & 700 - 1700     & 0.6 & 0.6 & 0.6 & 0.7 & 0.1 \\
   5000  & 1700 - 10000   & 0.3 & 0.3 & 0.3 & 0.3 & 0.1\\
\hline  
\end{tabular}
 \caption{{\small Results from the simulations of monochromatic photon 
parallel beams. 68$\%$ containment 
radius of the PSF in degrees.  Events for the different energy channels are selected 
on the basis of the reconstructed energy, as for real data.
}}\label{tab:monotabcr68}
\end{center}
\end{table}

\subsection{Monochromatic photons}
Monochromatic simulated data can be used for an ideal
characterization of the instrument response and were simulated for
different energies and different incident directions as previously
described. 

According to the definitions given in section \ref{sec:angres},
Table \ref{tab:monotabfwhm} and  \ref{tab:monotabcr68}  {\saim show the} values 
obtained for the
HWHM\footnote{Here we show the HWHM in place of the FWHM used in the figures in order 
to help for a direct comparison with the CR$_{68\%}$ and in the identification of the width of the 
radial profile (see e.g. Fig \ref{fig:crabsim_map1}, right panel).} 
(calculated from the PSF radial profile) and for the CR$_{68\%}$
(calculated from the PSF) for different
energies and different off-axis angles. We use both parameters
for symplifing the comparison
with previous papers, related both to gamma rays or to other wavelengths. 
{\saim In the rest of the paper we prefer adopting the FWHM to describe the 
angular resolution of the instruments in order to be more compliant with 
the definition in the multi-frequency domain.}

Note that for this
analysis we do not fit the PSF with any specific function, 
we just obtain the above parameters {\saim from} the raw 
distribution of the reconstructed directions (see e.g. Fig. 
\ref{fig:crabsim_map1} right panel). {\ss The 
source radial profile is given by the average source counts 
within circular crowns of increasing radii, {\saim centered at the source centroid.} 
The {\ssss instrumental} background 
counts are subtracted by evaluating their contribution in a circular 
crown at distance large enough to avoid contamination by the source tails
(see e.g. Fig. \ref{fig:agilecrabrealmap1}).} 
{\saim In Tables \ref{tab:monotabfwhm} and \ref{tab:monotabcr68} 
we also report the intrinsic error in the estimate, given by the 
bin size of the profile, which has a minimum value of $0.1^{\rm \circ}$. 
This value is the minimum bin size allowed for the generation of typical 
count maps in the AGILE real data for high significance sources. At the same time 
this value allowed us to have a statistical significance well above 5-$\sigma$ 
per bin in the radial profiles used to calculate the FWHM with the adopted number 
of simulated input events.}

{\saim Table \ref{tab:monotabfwhm} shows that the {\it AGILE} angular resolution, 
as inferred from the FWHM, is stable across the instrument field 
of view up to $30^{\rm \circ}$ of off-axis angle.
Table \ref{tab:monotabcr68} shows that, on the contrary, the CR$_{68\%}$ 
increases slightly with off-axis angle below 1000 MeV, since this parameter is much more affected by tails 
in the radial profile than the FWHM. This is expected, since at $30^{\rm \circ}$ off-axis angle tracks pass through much 
more material compared to the on-axis case: this causes a higher dispersion in the 
reconstructed direction due to the effect of multiple scattering, more pronounced below 1000 MeV 
(see Sec. \ref{sec:agilevsfermi}). 
}

\begin{figure}[!h]
\begin{minipage}[b]{0.45\linewidth}
 \includegraphics[width=9cm]{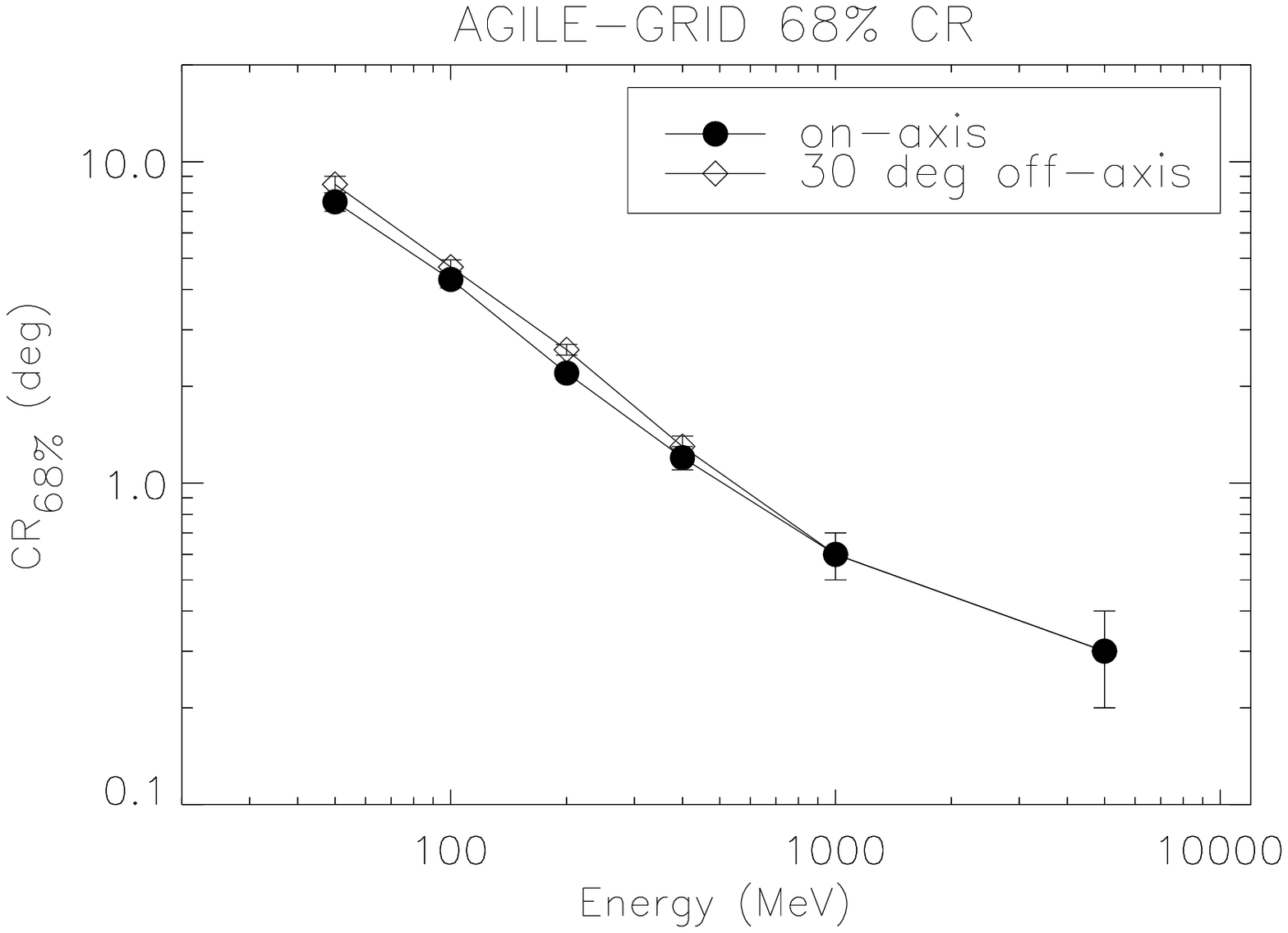}
\end{minipage}
\hspace{0.9cm}
\begin{minipage}[b]{0.45\linewidth}
 \includegraphics[width=9cm]{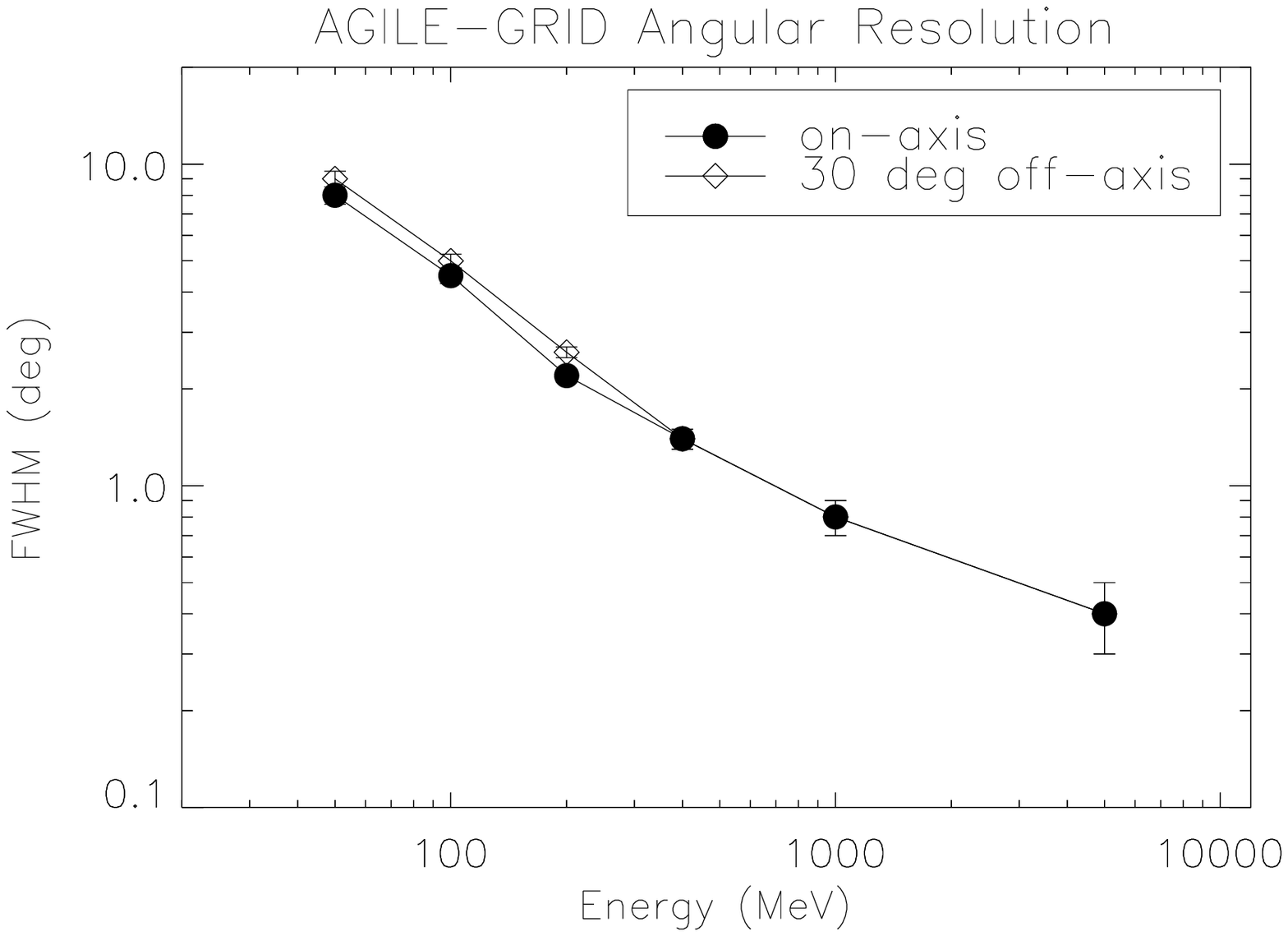}
\end{minipage}
\vspace*{-0.2cm}\caption{{\it Left panel:} AGILE-GRID 
68$\%$ containment radius versus photon energy for simulated monochromatic photons
of different incident angles. {\it Right panel:} angular resolution (FWHM) versus 
photon energy, shown for comparison to the CR68.}\label{fig:PSF-mono}
\end{figure}

\subsection{Simulations of Crab-like sources}
 
\begin{figure}[!h]
\begin{minipage}[b]{0.45\linewidth}
\centering
\includegraphics[height=6cm]{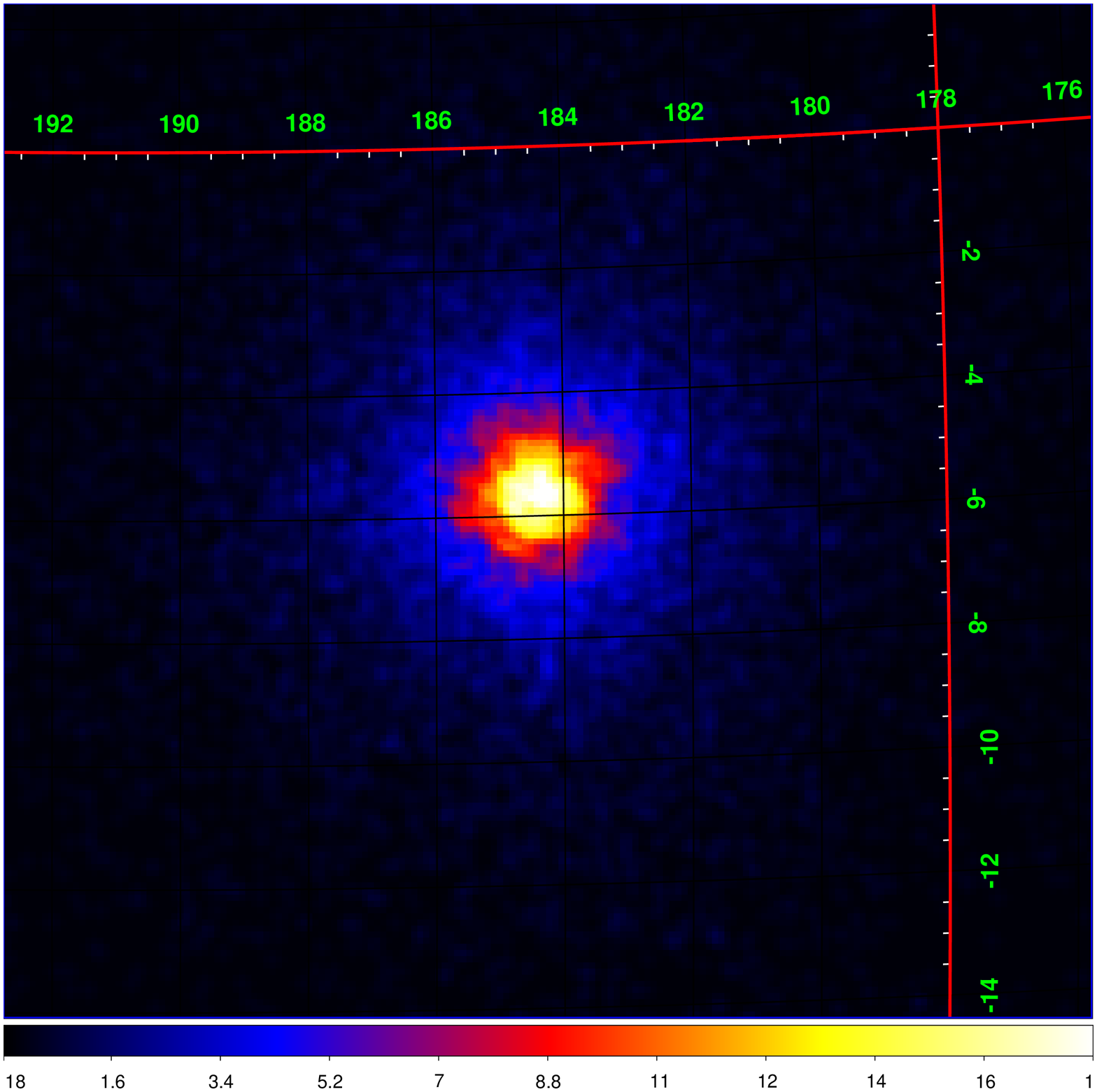}
\end{minipage}
\hspace{1cm}
\begin{minipage}[b]{0.45\linewidth}
\centering
\includegraphics[height=6.5cm]{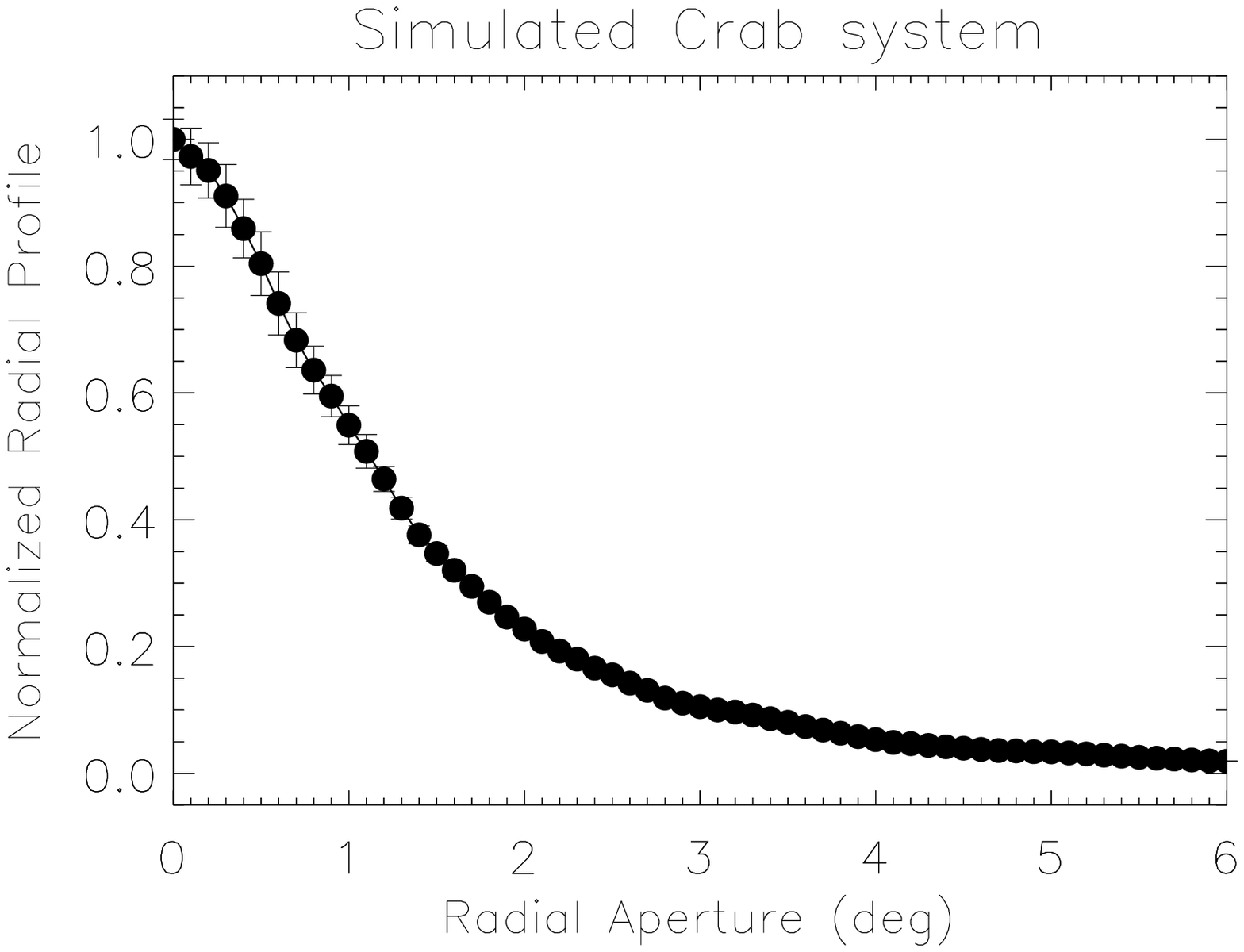}
\end{minipage}
\caption{{\it Left panel}: Count map of the simulated Crab system in the 100 - 400 MeV
 energy band.  Pixel size is $0.1^{\rm \circ}$. {\it Right panel}: {\ss 
Average count radial profile. The 
counts are normalized to the maximum value of the profile 
in order to reach 1 at $0^{\rm \circ}$. The error bars are the poissonian values.}}
\label{fig:crabsim_map1}
\end{figure}

Simulations of Crab-like spectrum sources were carried out as 
previously described and, as an example, Fig.
\ref{fig:crabsim_map1} shows a count map obtained from the MC
simulated data for the 100 - 400 MeV
energy band at the $(30^{\rm \circ},0^{\rm \circ})$ incident direction (left panel) and the
average radial profile of the source (right panel). The map was obtained by 
the simulated event list using the same pipeline as that for 
producing real data maps.

\begin{table}[!h]
\centerline{ \bf HWHM for the Crab simulations}
\begin{center}
\vspace{0.2cm}
\small 
\begin{tabular}{|c|c|c|c|c|c|c|c|c|}
  \hline
 Centroid  & Energy & $\theta=1^{\rm \circ}$ & $\theta=1^{\rm \circ}$ &
$\theta=30^{\rm \circ}$  & $\theta=30^{\rm \circ}$  & $\theta=50^{\rm \circ}$ & 
$\theta=50^{\rm \circ}$ & Err \\
  Energy &  Band  & $\phi=0^{\rm \circ}$   & $\phi=45^{\rm \circ}$ & 
 $\phi=0^{\rm \circ}$  &  $\phi=45^{\rm \circ}$   &  
$\phi=0^{\rm \circ}$  & $\phi=45^{\rm \circ}$& \\
(MeV) & (MeV) & (deg) & (deg) & (deg) & (deg) & (deg) & (deg) & (deg) \\[0.5ex]
  \hline

50          & 30 - 70      & 3.5  & 3.5 & 3.0 & 3.0 & 3.0 & 4.0 & 0.5 \\
100         & 70 - 140     & 2.0  & 2.0 & 2.0 & 2.0 & 2.4 & 2.4 & 0.4 \\
200         & 140 - 300    & 1.0  & 1.2 & 1.2 & 1.2 & 1.4 & 1.4 & 0.2 \\
400         & 300 - 700    & 0.6  & 0.6 & 0.8 & 0.8 & 0.8 & 0.8 & 0.2 \\
1000        & 700 - 1700   & 0.4  & 0.4 & 0.4 & 0.4 & 0.4 & 0.4 & 0.1 \\
5000        & 1700 - 10000 &0.3  & 0.3 & 0.3 & 0.3 & 0.3 & 0.3 & 0.1 \\
\hline
\hline
100 - 400   & 100 - 400 & 0.9  & 0.9 & 0.9 & 1.0 & 1.1 & 1.0 & 0.1 \\
400 - 1000  & 400 - 1000  & 0.45 & 0.45& 0.35& 0.35& 0.45& 0.35& 0.1  \\
100 - 50000 & 100 - 50000 & 0.50  & 0.50 & 0.40 & 0.40 & 0.45& 0.30 & 0.1 \\

\hline
\end{tabular}
 \end{center} 
\caption {HWHM of the {\it AGILE-GRID} PSF radial profile for different energy channels
and off-axis angles for the Crab simulations.}\label{tab:crabsim}

\end{table}

\begin{figure}[!h]
\begin{center}
{\includegraphics[width=13cm]{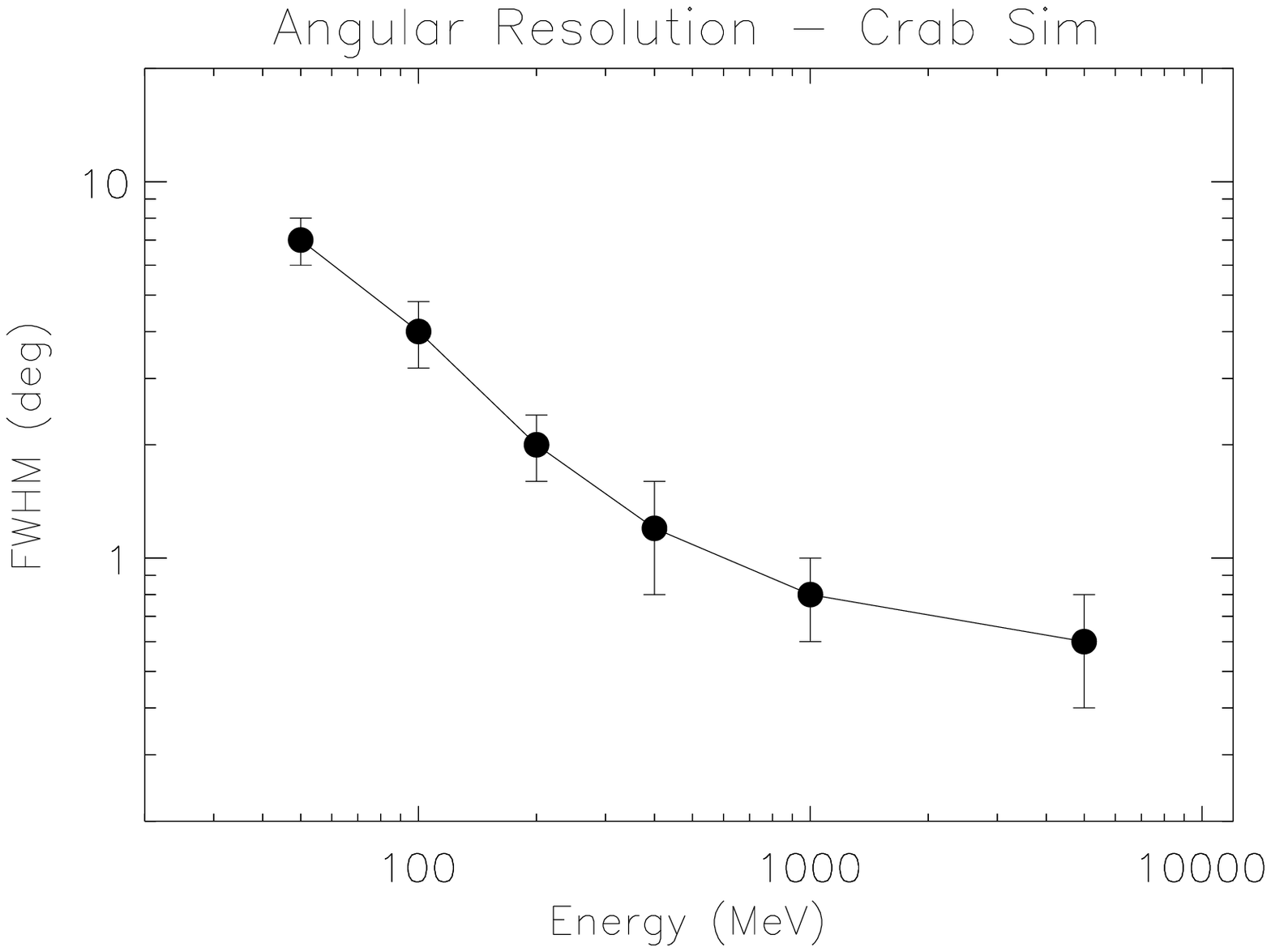}}
\end{center}  \caption{{\it AGILE-GRID} angular resolution versus photon 
energy for the simulation of an on-axis Crab-like source. }
  \label{fig:PSFCrabsim}
\end{figure}

\noindent Table  \ref{tab:crabsim} shows the results of the whole analysis
for the Crab simulations at varying photon energy and incident direction. 
{\ss The instrument 
response is stable within $30^{\rm \circ}$ across the FoV {\saim within the errors} and the
overall response is dominated by the zenith angle ($\theta$), since the 
dependency upon azymuthal angle is minimal, if any.}
Fig. \ref{fig:PSFCrabsim} shows the angular resolution as a function
of photon energy for on-axis incident directions. \\
\noindent We also report in Table \ref{tab:crabsim} the HWHM 
for the typical broad bands used in the AGILE-GRID data analysis, i.e.
the 100 - 400 MeV, the 400 - 1000 MeV channels and the 100 MeV - 50 GeV full band, 
which gives on average an angular resolution of $ 1.8^{\rm \circ} \pm 0.2^{\rm \circ} $, 
$0.8^{\rm \circ} \pm 0.2^{\rm \circ} $ and $0.9^{\rm \circ} \pm 0.2^{\rm \circ} $ for off-axis angles within $30^{\rm \circ}$
respectively. These values will be compared to the ones obtained with the in-flight data
analysis.  

{\saim It is interesting to compare the results of the Crab simulations to the 
monochromatic photon beams (see Table \ref{tab:crabsim} and \ref{tab:monotabfwhm}):
the values of the angular resolution per energy channels agree within the errors, showing that 
the contribution of possible energy channel cross talk is negligible.}

\section{In-flight data analysis}\label{sec:datacrab}

\begin{figure}[!h]
\begin{minipage}[b]{0.45\linewidth}
\centering
\includegraphics[height=6cm]{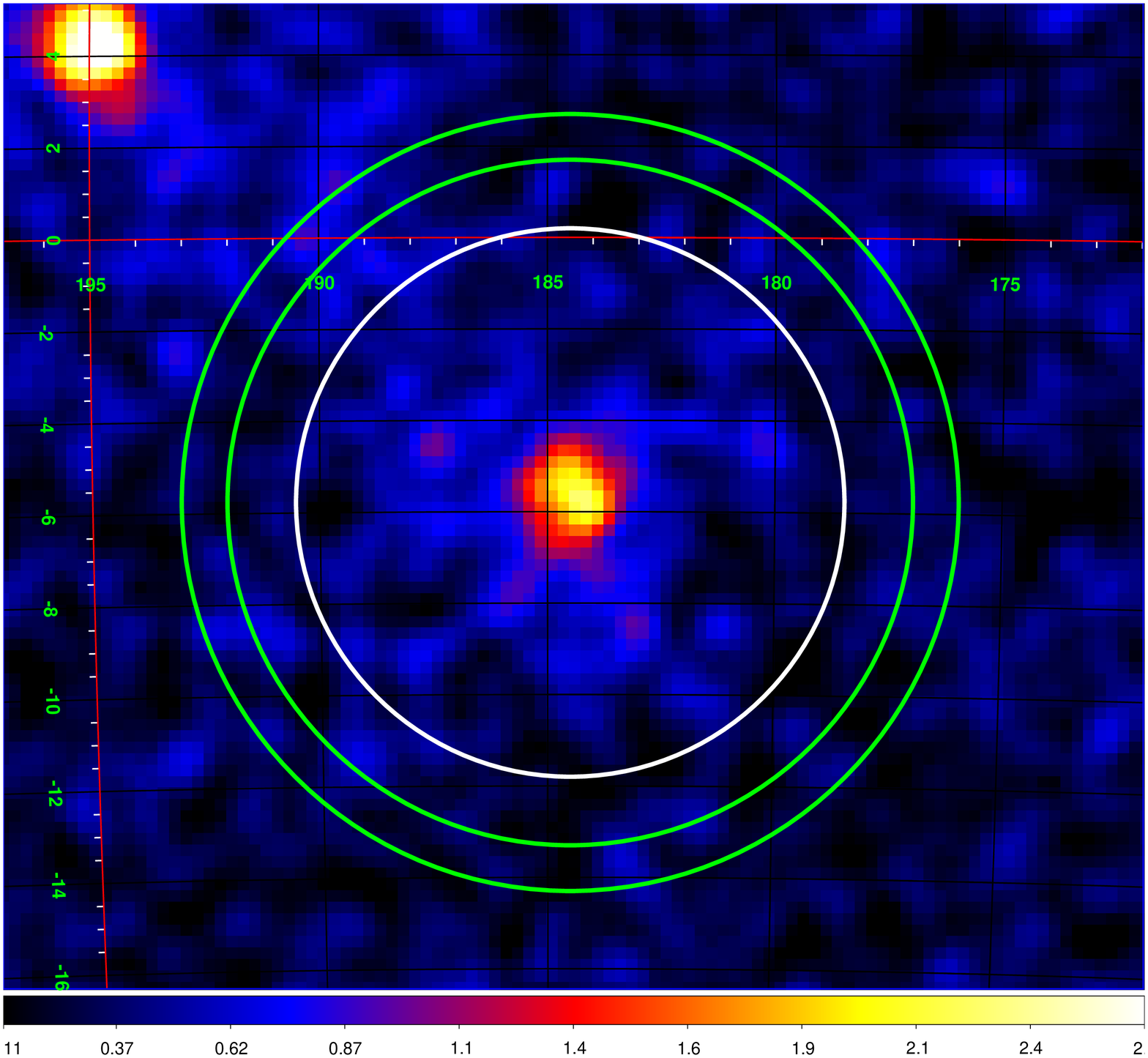}
\end{minipage}
\begin{minipage}[b]{0.45\linewidth}
\centering
\includegraphics[height=6.5cm]{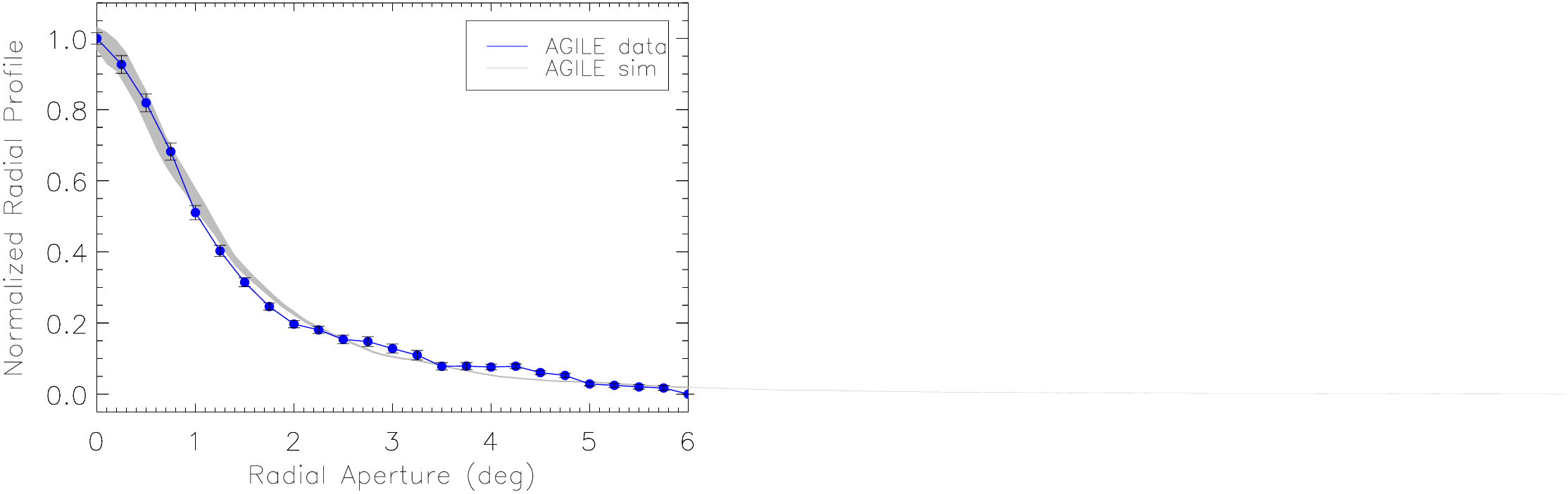}
\end{minipage}
\label{fig:agilecrabrealmap1}
\caption{{\it Left panel}: {\it AGILE-GRID} count map of the Crab system (pulsar $+$ Nebula)
for the period 2007-09-23T12:01:05 -
 in 2007-10-04T12:01:05 in the 100 - 400 MeV energy band; the white circle shows the maximum 
aperture ($6^{\rm \circ}$) used to produce the radial profile of the source shown in the right panel; 
green circles show
the crown region used to estimate the background emission 
($7.5^{\rm \circ}-8.5^{\rm \circ}$ radius).
{\it Right Panel}: Average count radial profile: blue data points are the in-flight data, while the 
gray shaded region shows the profile for the simulated crab system with errors. 
The counts are normalized to the maximum value of the profile 
in order to reach 1 at $0^{\rm \circ}$.
The source in the up left corner is Geminga.
}
\end{figure}

The {\saim whole simulation setup and the above results for the angular resolution}
can be validated by a comparison with the 
values obtained by in-flight data. The {\it AGILE} satellite has been operational since
July 2007 and the angular resolution of the {\it GRID} has been studied
both on-ground (Cattaneo et al., 2012; Cattaneo \& Rappoldi, 2013)
and in-flight (Chen at el., 2013). Here however we focus in particular on
the case of the Crab source (pulsar + Nebula), whose spectral energy
distribution in the gamma-ray energy band is dominated by the
pulsar and is described by a power law with a spectral index of $\sim$2.1.
{\ss As previously mentioned, this spectral energy distribution
is representative of the majority of the gamma-ray sources and
we therefore think that it is  an ideal test case to assess the instrument response,
also in view of future gamma-ray missions under study (see next section).}
We selected {\it AGILE}  data from a long-term pointing of the region (11
days, 2007-09-23T12:01:05 - 2007-10-04T12:01:05), during which
the source was located at an off-axis angle $\le 30^{\rm \circ}$. {\ss Due
to the resulting overall exposure of the region, we focus hereafter
on two wide energy bands in order to obtain a good photon statistics for a robust
assessment of the angular resolution: the 100 - 400 MeV and
400 MeV - 1 GeV energy bands.} Fig. \ref{fig:agilecrabrealmap1} and 6
show the Crab intensity map for the two
energy bands respectively and the average count radial profile for the source. The
FWHM associated to these profiles turns out to
be $ 2.5^{\rm \circ} \pm 0.5^{\rm \circ}$ in the 100 - 400 MeV energy range, and 
$ 1.2^{\rm \circ} \pm 0.5^{\rm \circ}$ for the 400 MeV
- 1 GeV band. The good agreement with the results from the simulations 
(see Table \ref{tab:crabsim}) validate the whole setup and 
data analysis carried out for the simulations.

\begin{figure}[!h]
\begin{minipage}[b]{0.45\linewidth}
\centering
\includegraphics[height=6cm]{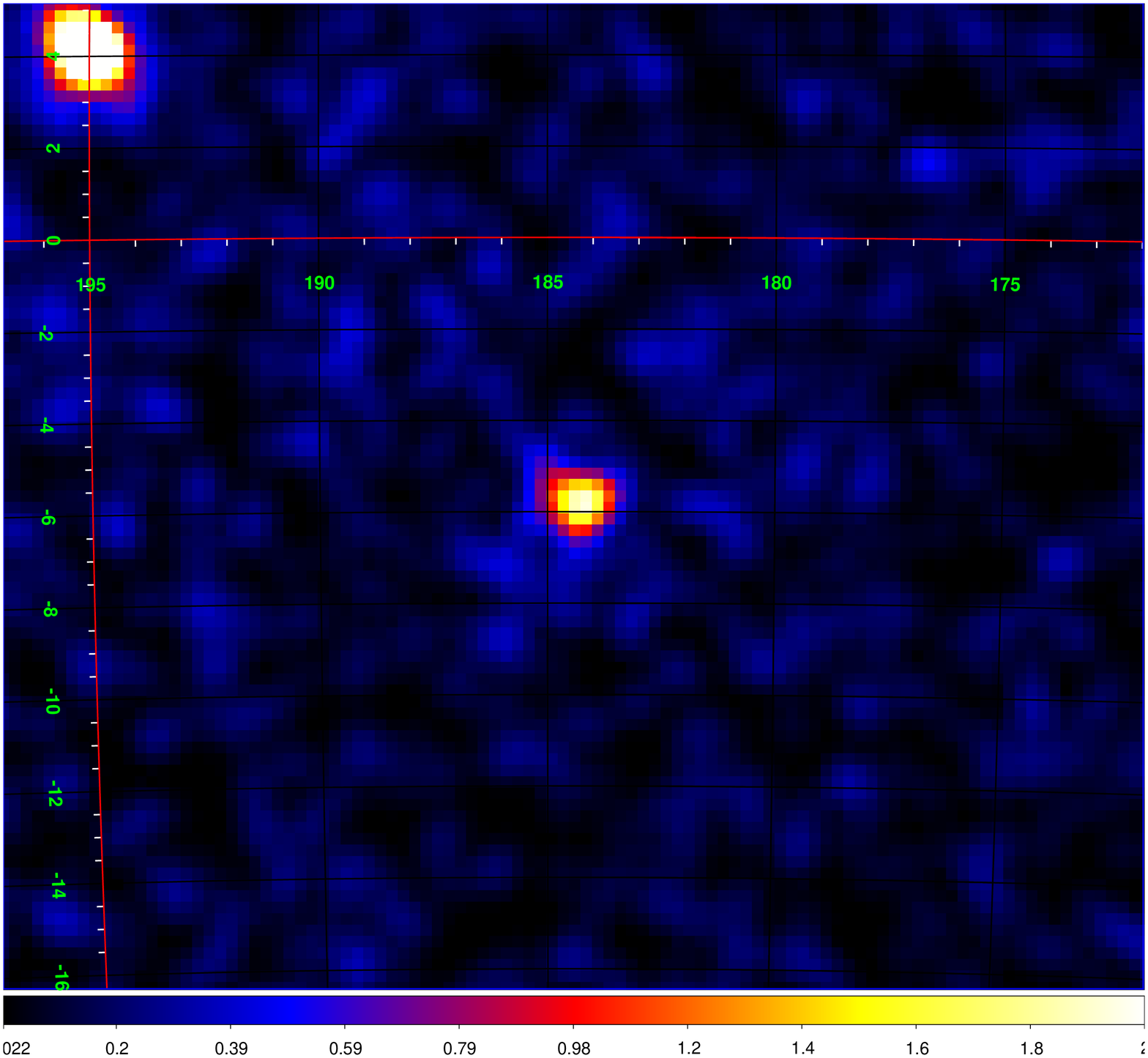}
\end{minipage}
\begin{minipage}[b]{0.45\linewidth}
\centering
\includegraphics[height=6.5cm]{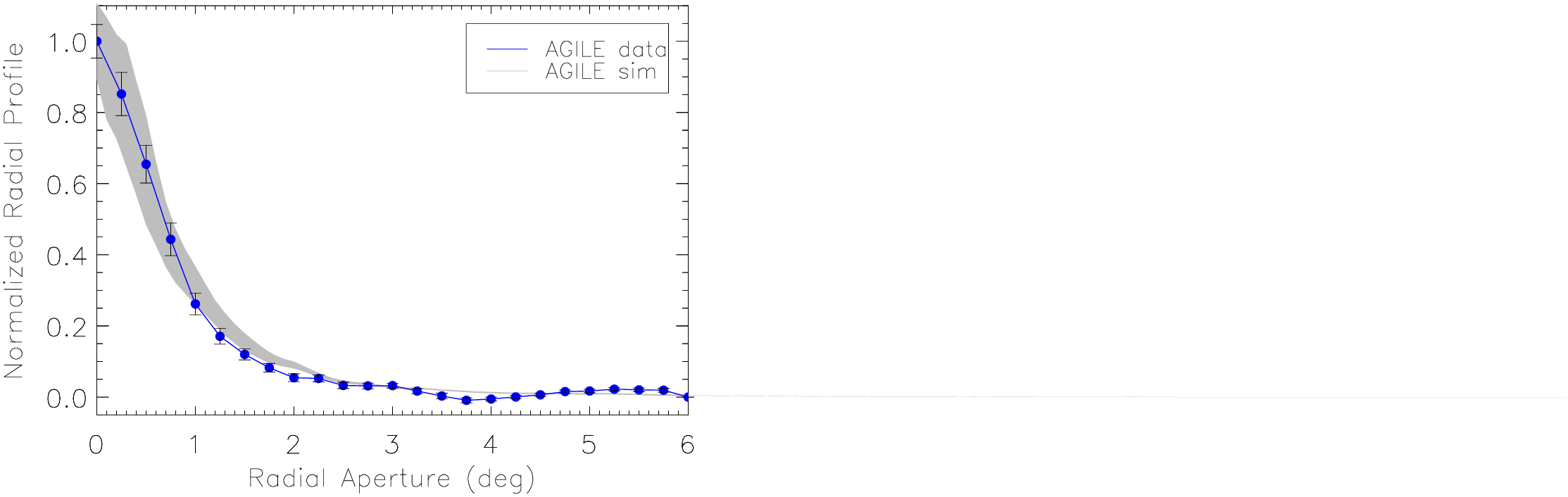}
\end{minipage}
\label{fig:agilemap2}
\caption{{\it Left panel}: {\it AGILE-GRID} count map of the Crab system (pulsar $+$ Nebula)
in the 400 - 1000 MeV energy band.
{\it Right Panel}: Average count radial profile: blue data points are in-flight data, while the 
gray shaded region shows the profile of the simulated crab system with errors. The counts are normalized to the maximum value of the profile 
in order to reach 1 at $0^{\rm \circ}$.
}
\end{figure}

\section{AGILE and \textit{Fermi}: comparison of in-flight data for the Crab system}
\label{sec:agilevsfermi}

As already mentioned in section \ref{sec:intro}, 
although implementing the same silicon-tungsten detector concept,
the instrument configurations for the {\it AGILE-GRID} and {\it Fermi-LAT}
detectors are quite different in terms of converter thickness,
distance between trays and spacing between adjacent silicon
strips (see Table A1). In the case of the {\it AGILE-GRID},
the Silicon Tracker is composed of 10 tungsten converter planes of
homogeneous thickness (0.07 radiation length each) plus 2
additional planes without converter (for a total of 12 planes);
the overall GRID radiation length is 
$\sim 1$X$_{\rm \circ}$ \citep{barbiellini02,prest03}.
 The {\it Fermi-LAT}  has 12 tungsten planes of $\sim 0.03$
radiation length thickness (``LAT-front''), and an additional set of 4
tungsten planes of 0.18 radiation length (``LAT-back'') plus 
two planes without converter \cite{atwood09}.
In both instruments, each tungsten plane is interleaved with 2 
layers of silicon strip detectors which are sensitive to charged
particles and are used for the x,y positioning of the
e$^{+}$/e$^{-}$ pair track resulting from the pair conversion. The
ratio of strip pitch to vertical spacing between tracker planes is
0.007 for both {\it Fermi-LAT} and {\it AGILE-GRID}.

\begin{figure}[!h]
\begin{center}
{\includegraphics[width=10cm, angle=270]{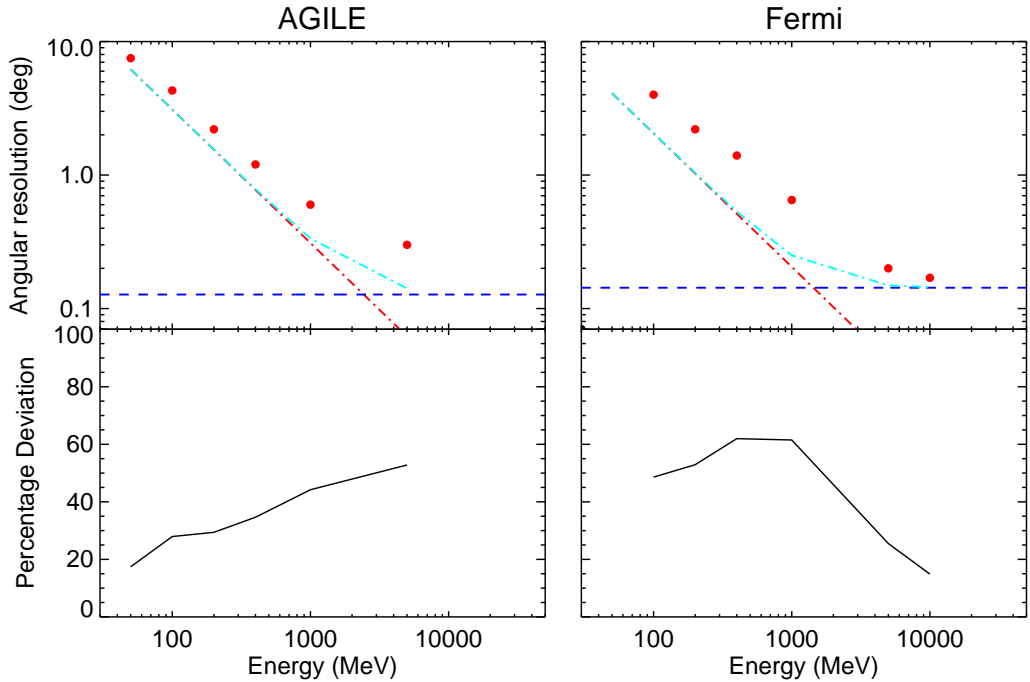}}
\end{center}  
\caption{
Comparison of the expected limiting values for the angular
resolution (dot-dashed cyan line) to the measured ones (red points). {\it Left panel: AGILE-GRID} simulated Crab data;
{\it right panel: Fermi-LAT} PSF68 
(\url{http://www.slac.stanford.edu/exp/glast/groups/canda/lat_Performance.htm}).
{\it Top panels}. The overall limiting angular resolution is 
plotted in cyan and is
obtained by the combination of: 1. the ratio of the spatial resolution to the 
distance of consecutive trays (dashed blue line); 2. the multiple scattering limit 
(red dot-dashed line). {\it Bottom panels}. Percentage Deviation of the 
measured angular resolution from the overall expected limiting value. 
}

  \label{fig:multiplescattering}
\end{figure}

\begin{figure}[ht]
\begin{center}

    \begin{minipage}[b]{0.45\linewidth}
       \centering
       \includegraphics[width=6cm]{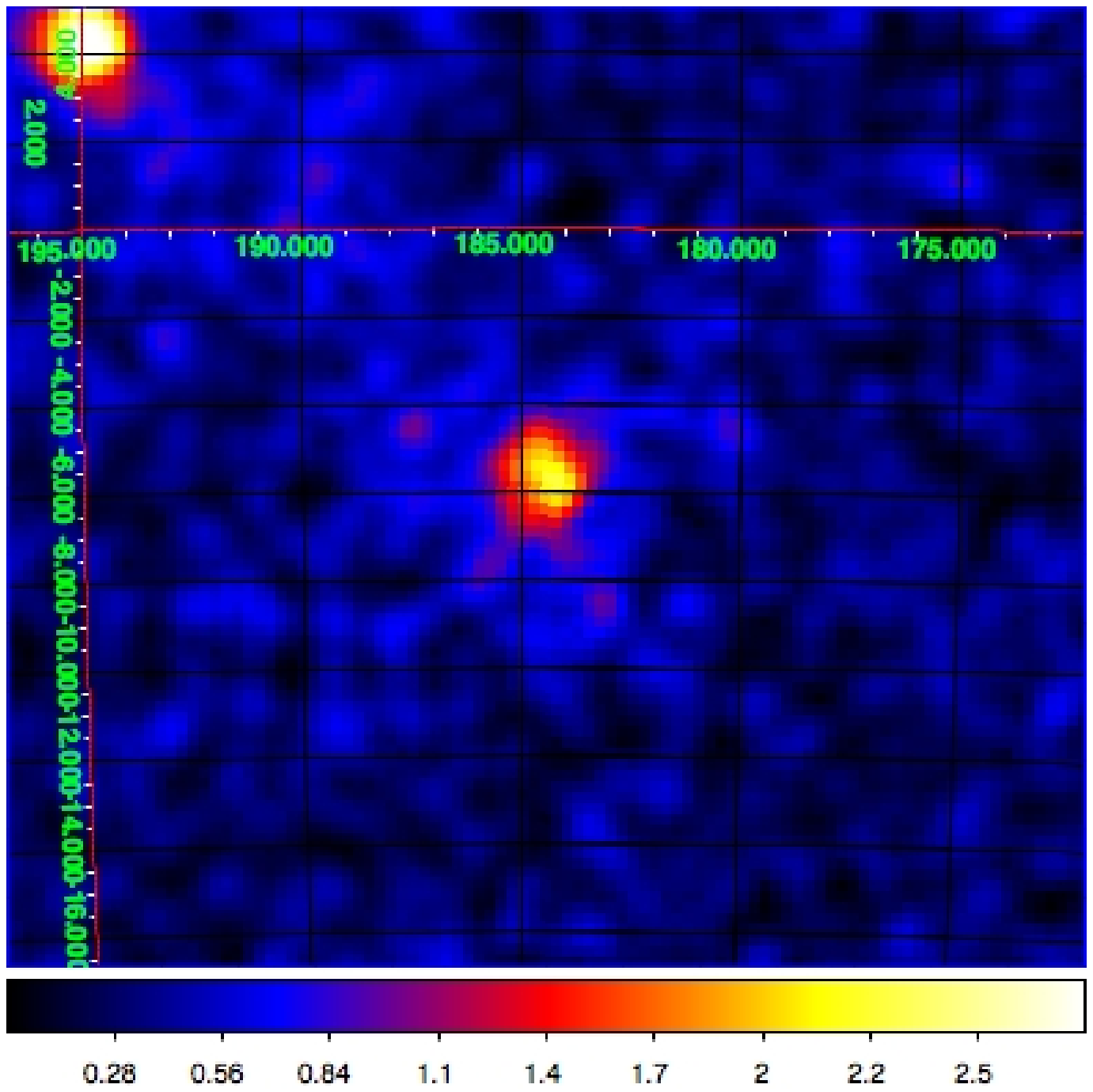}
    \end{minipage}
 \hspace{0.1cm}
    \begin{minipage}[b]{0.45\linewidth}
       \centering
       \includegraphics[width=6cm]{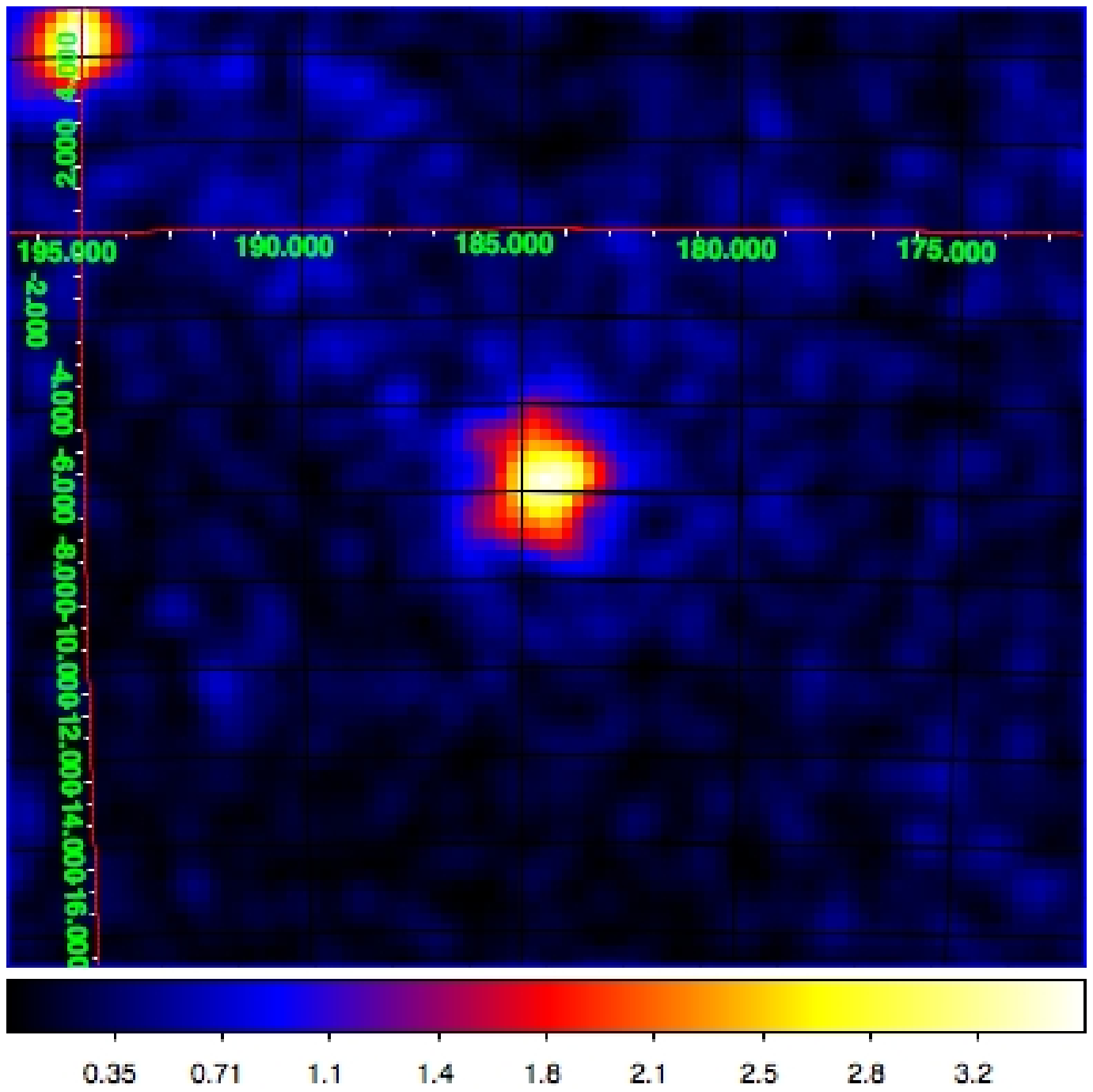}
    \end{minipage}

   \begin{minipage}[b]{0.45\linewidth}
       \centering
       \includegraphics[width=6cm]{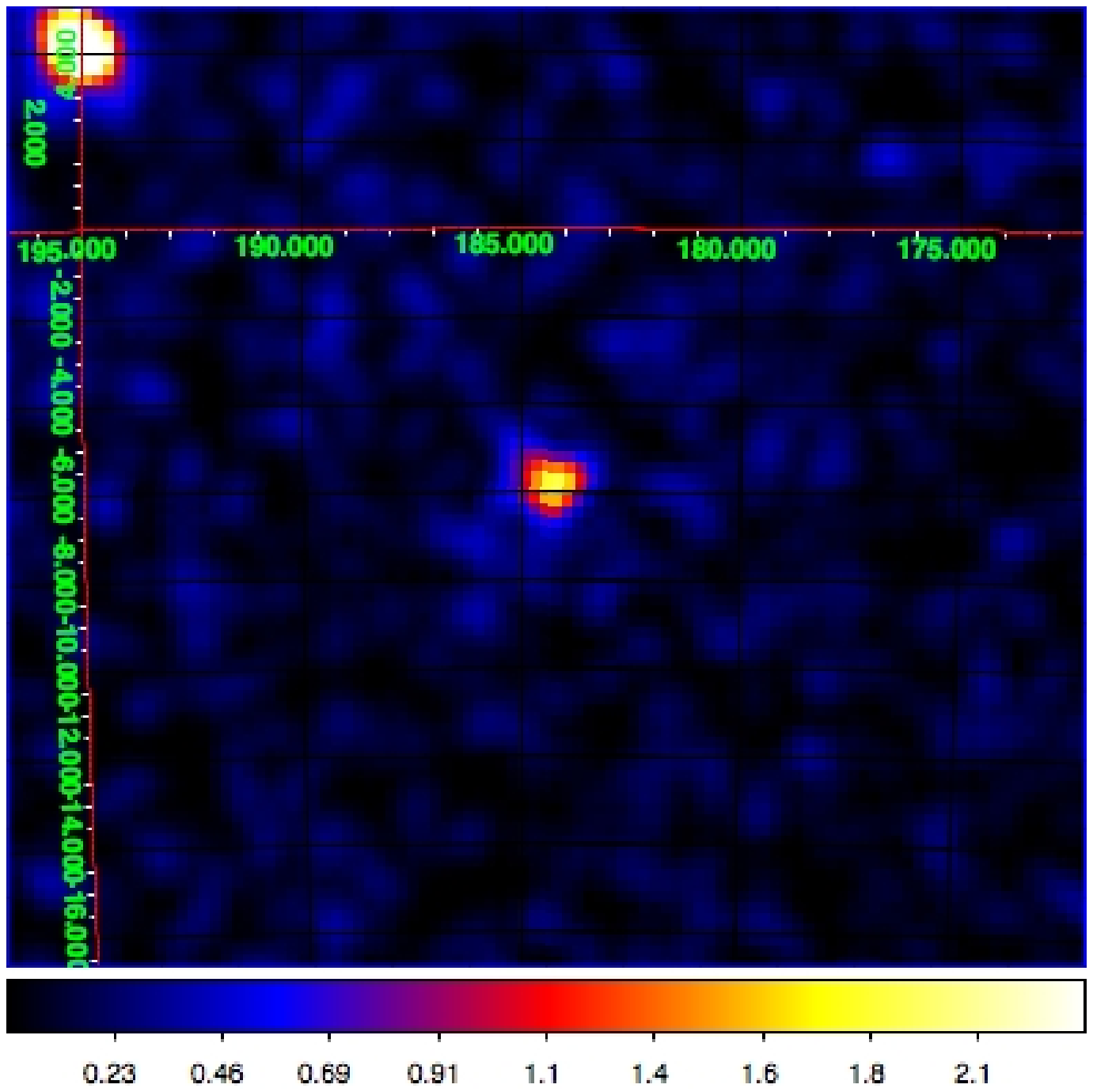}
    \end{minipage}
 \hspace{0.1cm}
    \begin{minipage}[b]{0.45\linewidth}
       \centering
      \includegraphics[width=6cm]{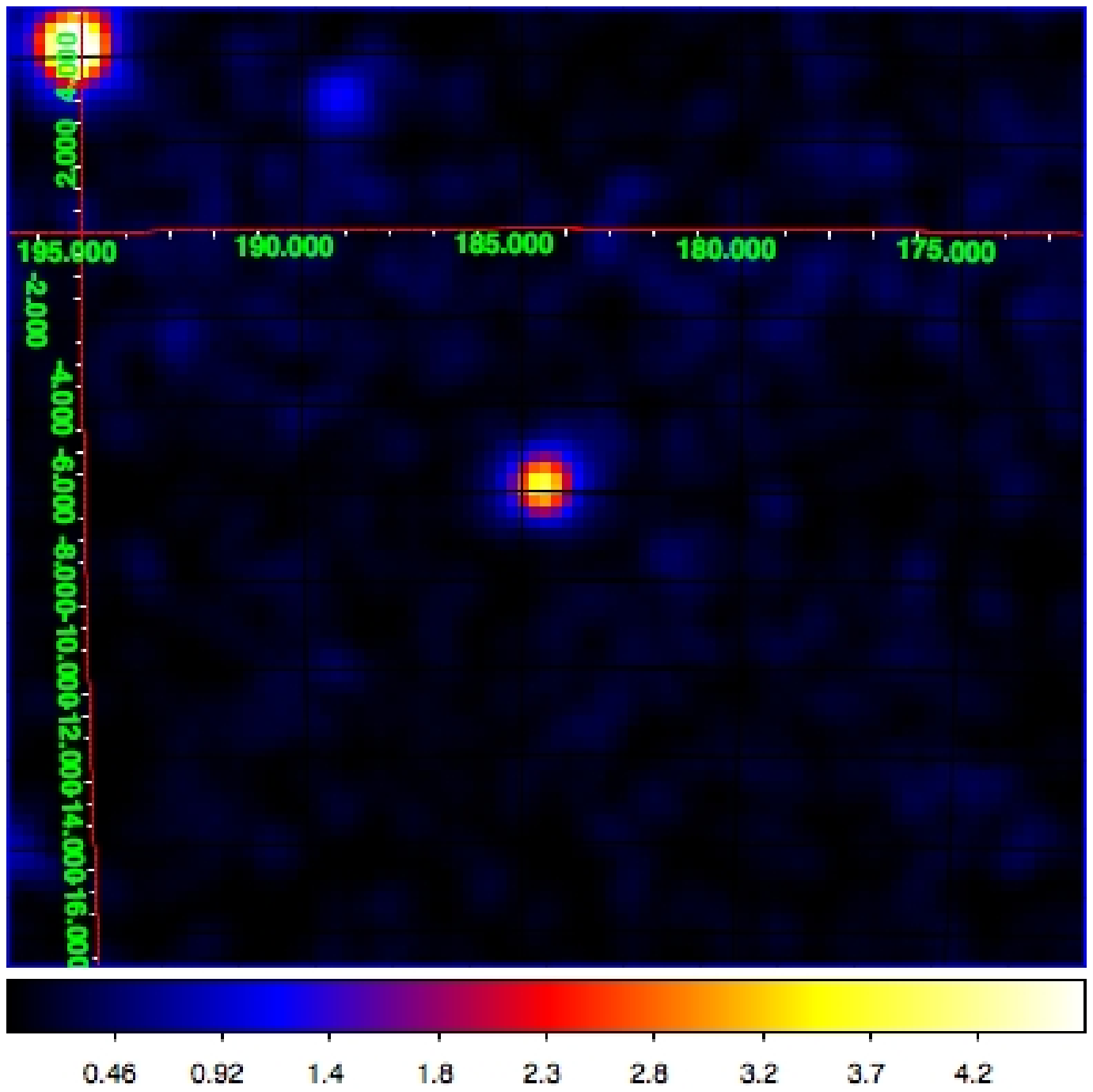}
    \end{minipage}

 \caption{{\it AGILE-GRID} (left panels) and {\it Fermi-LAT} (right panels) 
in-flight count maps for the Crab
region in the 100 - 400 MeV and 400 - 1000 MeV energy bands
respectively. Pixel size is $0.25^{\rm \circ}$. {\ssss Note that the color intensity scale is set so that
each map has the maximum value of the scale corresponding to the Crab peak counts.} }
\label{fig:agilefermimappe}
 \end{center}
\end{figure}

\begin{figure}[ht]
\begin{center}
\begin{minipage}[b]{0.45\linewidth}
\centering
\includegraphics[height=6.cm]{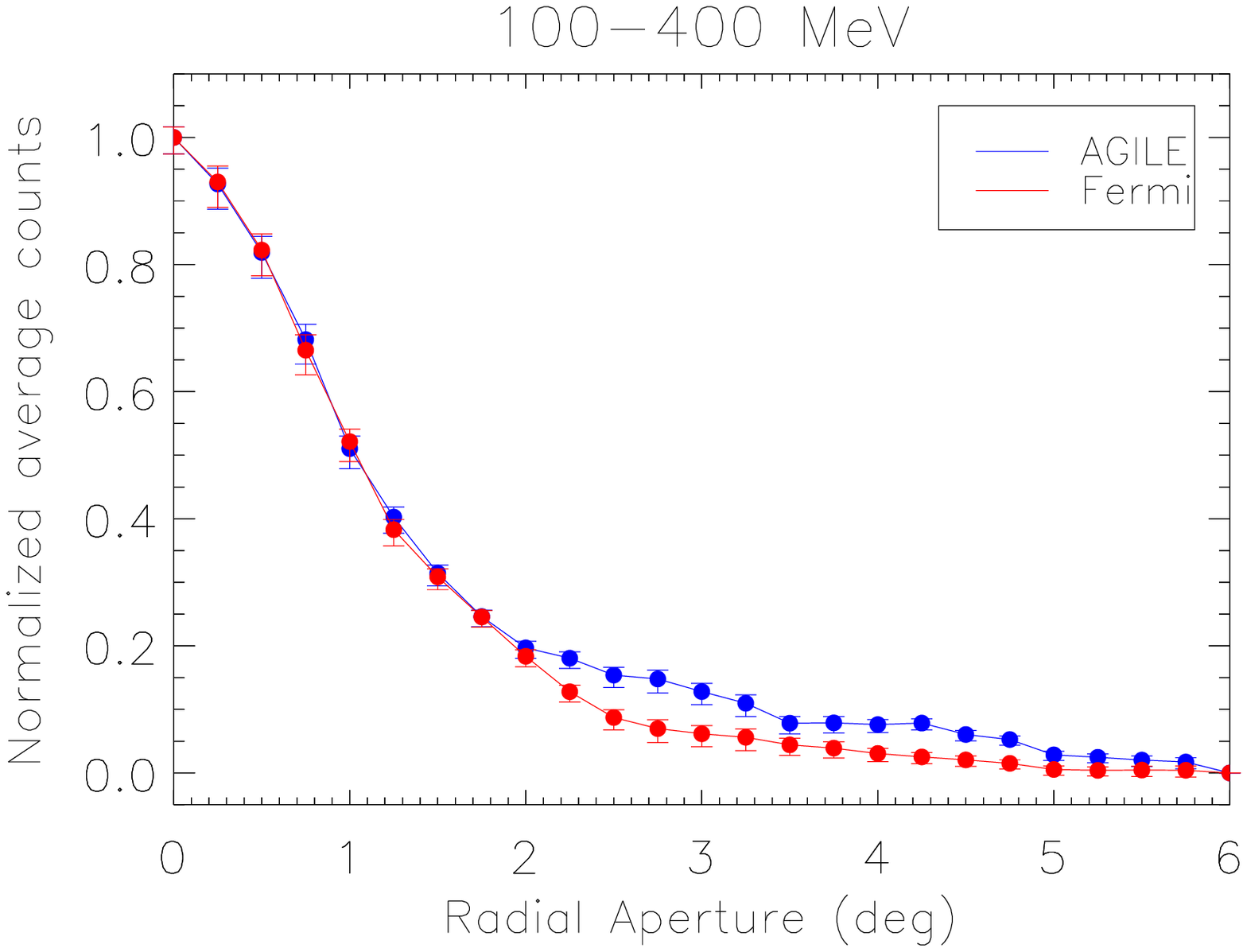}
\end{minipage}
\hspace{1cm}
\begin{minipage}[b]{0.45\linewidth}
\centering
\includegraphics[height=6.cm]{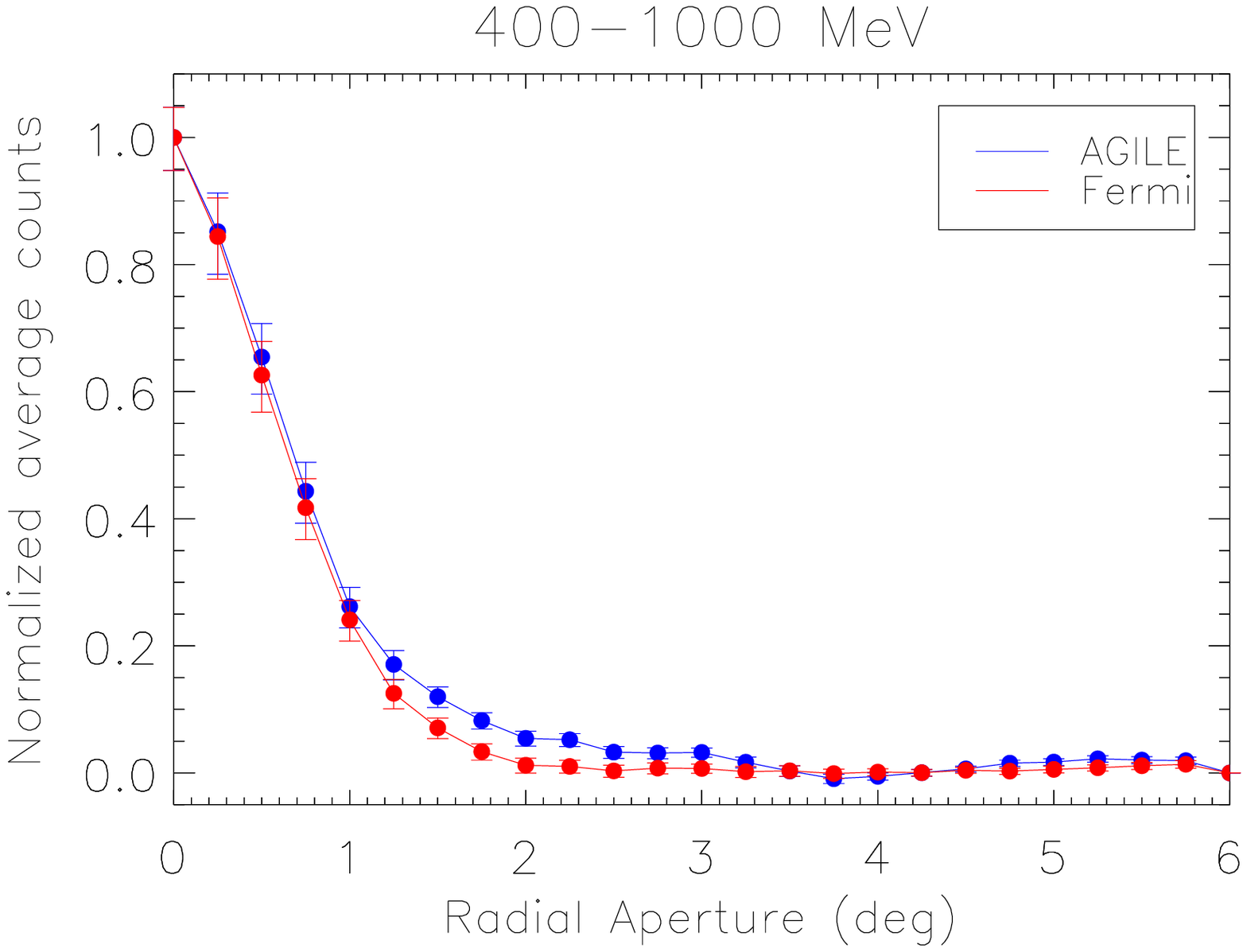}
\end{minipage}
\vspace*{-0.2cm} 

\caption{Average count radial
profiles for circular aperture of increasing radii at steps of $0.25^{\rm \circ}$
 of the Crab (pulsar + Nebula); in-flight data for the
{\it AGILE-GRID} (blue data points) and {\it Fermi-LAT} (red data
points). {\it Left panel:} 100 - 400 MeV energy range; {\it right panel:} 
400 - 1000 MeV energy range. }
\label{fig:radprofcomparison} 
\end{center}
\end{figure}

{\saim An important
difference between the two instruments,} is
the Silicon tracker readout system, {\ss analog for {\it AGILE-GRID} and 
digital for {\it Fermi-LAT}. Barbiellini et al. (2002) and Prest et al. 
(2003) 
describe
extensively the {\it GRID} dedicated front-end electronics, {\saim
characterizing the effective spatial resolution for particle
incidence between $0^{\rm \circ}$ and $30^{\rm \circ}$ and comparing the performances 
of digital to analog readout.} Although the {\it GRID} analog 
readout is structured to read only odd-numbered
strips with no signal pick up at even-numbered strips (``floating
strip readout''), the capacitive coupling between adjacent strips
allows to obtain a complete sampling of the particle hit positions,
discriminating between hits involving directly read and not-read
microstrips. Typically, each particle hit is characterized by a
signal spread out over several adjacent readout-strips (a ``cluster''). 
{\it GRID} events produce energy deposition
histograms per cluster which are very well characterized and
typically result in 2-3 triggered strips depending on off-axis angle  
(Barbiellini et al. 2002; Fedel et al. 2000).
 Note that also for on-axis incidence a cluster is composed of a minimum of 
2 strips. The effective spatial
resolution obtained by the {\it GRID} for particle incidence between
$0^{\rm \circ}$ and $30^{\rm \circ}$ is $ \delta_s \sim 40 \, \mu \rm m$,
substantially less than the Silicon microstrip size of 121 $\mu \rm m$
\cite[see Fig. 15 in ][]{barbiellini02}. 
The advantage of the analog vs. digital readout for the resulting spatial resolution 
of silicon strip detectors is discussed in \cite{barbiellini02}  for the AGILE tracker 
configuration, showing an improvement of 
a factor of 2 (i.e. $ \delta_s \sim 40 \, \mu \rm m$
vs. $ \delta_s \sim 80 \, \mu \rm m$). 

The {\it Fermi-LAT} system is based on a digital
readout. 
The LAT Tracker is non-homogeneous and is characterized by two different 
values for the angular resolution, the
one for the ``LAT-front'' and the of the ``LAT-back'' \citep{ackermann13}. 
In the following, we use only LAT-front data for comparison with the
{\it AGILE-GRID}, proving the best angular resolution for \textit{Fermi}. 
Note also that \textit{Fermi} usually
operates in scanning mode, with sources observed most of the time at large 
off-axis angles, with a consequent degradation of the angular resolution 
for a given source, {\saim compared to the on-axis performance}. 

In order to compare the angular resolution 
of the two instruments at similar conditions, we selected \textit{Fermi} data 
of the Crab system in pointing mode. We analyzed 
therefore a comparable set of data, with an exposure ensuring a similar photon 
statistics and with both gamma-ray detectors in pointing mode. 
The {\it AGILE} data in pointing mode\footnote{An analysis of
AGILE-GRID data in ``spinning mode'' will be discussed in a
forthcoming paper by Lucarelli et al., 2015.} used for this comparison are the ones
described in section \ref{sec:datacrab}. The {\it Fermi} data
consist of 4 days of integration in the period 2012-07-04T23:24:44
- 2012-07-08T10:44:43 during which the satellite was stably
pointing at the Crab following a gamma-ray flaring episode, {\ss with the
source located at $10^{\rm \circ}$ of off-axis angle in the FoV. 
We used the best available quality cut for the Fermi data,
(the``ULTRA-CLEAN'' event class, {\ssss LAT-front photons}, P7REP) in order to optimize
the angular resolution of the data.} We
focus on the analysis of two representative energy bands for {\it AGILE} in order to
have a robust statistics for both instruments:  100 - 400
MeV, and 400 - 1000 MeV.

Fig. \ref{fig:agilefermimappe}
 shows the intensity maps for the two gamma-ray
telescopes in the two energy bands and Fig.  \ref{fig:radprofcomparison}
shows the {\saim corresponding} radial profiles. The resulting
angular resolution deduced from the in-flight data are
the same, even though the {\it AGILE} data show a more pronounced 
non-gaussian tail, due to the converter thickness: the FWHM are
$2.5^{\rm \circ} \pm 0.5^{\rm \circ}$ and $1.2^{\rm \circ} \pm 0.5^{\rm \circ}$ in the energy range 100 - 400 MeV and 
400 MeV - 1 GeV respectively. 

As already mentioned, several factors determine the instrument angular 
resolution as a function of
gamma-ray photon energy, including: \textit{(1)} multiple
scattering, \textit{(2)} the effective spatial resolution,
\textit{(3)} the photon energy reconstruction, \textit{(4)}
specific particle track reconstruction algorithms, \textit{(5)}
quality cuts. 
{\saim Fig. \ref{fig:multiplescattering} shows the relative contibution of 
these factors as a function of energy for both trackers, {\it AGILE} and 
{\it Fermi}.
Regarding the multiple scattering, we assume an effective radiation length
per tray of 0.085 and 0.04 for {\it AGILE-GRID} and {\it Fermi-LAT} respectively, taking 
into account the contributions of the Tungsten converter and the supporting 
material\footnote{{\saim In the case of {\it AGILE} the effective radiation length due to the
supporting material for each tray is $\sim 0.015 X_{\rm \circ}$
from laboratory measurements; since {\it Fermi} has a similar tray structure 
\cite{atwood09} we assume a similar contribution of $\sim 0.01 X_{\rm \circ}$.} }. {\saim Top panels show that
the multiple scattering effect dominates the overall angular resolution
up 700 MeV and 350 MeV respectively for the two instruments. Bottom panels
show the percentage deviation of the measured angular resolution to the 
overall expected limiting value. The amplitude of the deviation quantifies 
the reconstruction accuracy, that is optimized in different energy ranges for the
two instruments: the accuracy is below 40$\%$ 
up to 400 MeV in the case of {\it AGILE} and above 2 GeV in the case of {\it Fermi}.} 
}
Besides the different tracker configuration, 
crucial features distinguishing the two instruments and affecting the final 
reconstruction accuracy are the effective spatial resolution and the quality
of the reconstruction algorithms\footnote{Note that ongoing developments for 
the track reconstruction algorithms and events classification for both {\it AGILE} 
(updated FM filter)  and {\it Fermi-LAT}  (PASS8) may lead to further 
improvements of the current angular resolution.}. Therefore we believe that 
in view of future gamma-ray missions 
\cite[e.g.][]{morselliglight,galper13} the optimization of the angular resolution
should not only rely on a reduced converter thickness, but also on the implementation
of an analog readout system {\ssss if the power budget allows it}, together with 
an optimization of the reconstruction algorithms. } 


\section{Conclusions}

The current generation of gamma-ray space instruments is based on
Silicon detector technology and associated electronics. Both {\it AGILE-GRID}
and {\it Fermi-LAT} show a quite stable performance in orbit which is the
basis for prolonged operations ({\it AGILE} is at its 9th year of
life in space, {\it Fermi} at its 8th year). Compared to the
previous generation, the instrument performance improvements both
in terms of sensitivity and angular resolution is well
established.

In this paper, we summarized the main results concerning the {\it GRID}
angular resolution, a crucial feature of the {\it AGILE} instrument, at the
basis of the scientific performance of a gamma-ray detector
together with its FoV and background rejection capabilities {\saim at energies 
below 400 MeV, allowing for the best exploitation of the instrument configuration. We
showed} that the {\it AGILE-GRID} angular resolution is optimized given the overall
characteristics of the detector and  
allows for state-of-the-art mapping of Galactic and extragalactic
regions/sources: the FWHM for
off-axis angles in the range $0^{\rm \circ} - 30^{\rm \circ}$ is $\sim 4^{\rm \circ}$ at
100 MeV, and $\sim 0.8^{\rm \circ}$ at 1 GeV. The angular resolution is
quite uniform in the FoV up to $30^{\rm \circ}$ off-axis. The FWHM
 obtained from in-flight data in pointing mode
 is $2.5^{\rm \circ}$ in the range 100 - 400 MeV, and $\sim 1.2^{\rm \circ}$
 in the 400 MeV - 1 GeV. 

Although the {\it AGILE-GRID} multiple scattering
due to the heavy converter is relatively high, we proved that a 
crucial role is played by the optimization of the readout
system (analog) of the Silicon Tracker and of the particle track
reconstruction algorithms\footnote{It is interesting to note that the
 AGILE Tracker configuration is quite similar to the basic
element of the gamma-ray instrument currently under study for the
GAMMA-400 mission (Galper etal. 2013). The analog readout of a
Silicon Tracker with AGILE-like characteristics is required to
optimize the angular resolution with a thick converter (that in the case of
GAMMA-400 is currently designed to be $\sim 0.08 \, X_{\rm \circ}$ per
plane).}.

The {\it GRID} angular resolution as a function of gamma-ray energies is shown in 
Fig.  \ref{fig:PSF-mono} as resulting from simulations. These values are in good agreement
with the ones deduced from in-flight data as demonstrated in this
paper for Crab-like sources. 

Furthermore, by a direct comparison of in-flight data of the Crab
system, we find that despite the differences in
structure, geometry, readout system and overall size, the
{\it AGILE-GRID} and {\it Fermi-LAT} front show similar angular resolutions at energies
between 100 MeV and 1 GeV, due to different optimizations of the readout system and 
reconstruction algorithms. 

\section*{Acknowledgements}
We ackowledge several discussions with our colleagues of the Fermi Team. The \textit{AGILE} mission is funded by the Italian Space Institute (ASI) 
with scientific and programmatic participation by the Italian Istitute of Astrophysics 
(INAF) and the Italian Insitute of Nuclear Physics (INFN). Our research is partially supported
by the ASI grants I/042/10/0, I/028/12/0 and I/028/12/02. We would like to thank
 the referee for the careful review and for providing valuable comments that helped 
in improving the contents of this paper. 


\newpage

\section{Appendix: The AGILE GRID vs. Fermi-LAT}


AGILE is an ASI Small Scientific Mission (Tavani et al. 2009) of
total weight of 320 kg. It carries a scientific instrument
dedicated to high-energy astrophysics whose heart is the Gamma-Ray
Imaging Detector (GRID). Complementary items are the imaging
Super-AGILE detector sensitive in the range 20-60 keV (Feroci et
al., 2007), the Mini-Calorimeter (Labanti et al. 2006), and the
Anticoincidence system (Perotti et al. 2006). AGILE was launched
in April 2007 and is operational in an equatorial orbit of average
height of 530 km.

\textit{Fermi} is a NASA mission of a large class with a broad
international collaboration. Its imaging gamma-ray instrument is
based on a 16-unit Tracker (LAT) (Atwood et al., 2009), which is
complemented by a massive Calorimeter and an Anticoindence system.
Fermi was launched in June 2008 in an orbit with
inclination of $25^{\rm \circ}$ and average height of 550-600 km.

Table A1 summarizes the main parameters of the two instrument configurations,
which are relevant for the angular resolution determination.

\bigskip

\begin{table*}[!h]
\centerline{\bf Table A1: A comparison between the AGILE-GRID and
Fermi-LAT} \vspace*{0.2cm}
\begin{tabular}{|l|c|c|}
  \hline
  Parameter & AGILE-GRID & Fermi-LAT \\
  \hline
  Number of towers              & 1 & 16 \\
  Total number of Tracker planes & 12 & 18 \\
  Vertical spacing ($ s $) between adjacent planes & 1.8 cm & 3.2 cm\\
  Silicon tile size     & 9.5 x 9.5 $\rm cm^2$ & 8.95 x 8.95 $\rm cm^2$ \\
  Silicon detector array for each plane & 4 x 4 & 4 x 4 \\
  Silicon-strip pitch ($\delta_P$) & 121 $\rm \mu m$ & 228 $\rm \mu m$ \\
  Readout pitch  & 242 $\mu m$& 228 $\rm \mu m$ \\
  Signal readout & analog & digital \\
  Ratio $\delta_P / s$     & 0.007 &  0.007 \\
  Tungsten converter thickness per plane & 0.07 $X_o$ & 0.03 $X_o$  (front)\\
   & &  0.18 (back) \\
  Number of planes with W converter  & 10 & 12 (front) \\
                                      &    & 4 (back) \\
On-axis total radiation length        & 0.9 & 0.5 (front) \\
                                      &     & 0.8 (back) \\
Total n. of readout channels & 36,864      & 884,736   \\
Power consumption/channels   &  400 $\mu$W     &  180 $\mu$W   \\
 \hline
\end{tabular}

\vspace{0.2cm}
\end{table*}
\end{document}